\documentclass[namedreferences]{SolarPhysics}

\usepackage[optionalrh,solaenum]{spr-sola-addons} 
\usepackage{subfig}
\usepackage{caption}
\usepackage{graphicx}        
\usepackage{color}           
\usepackage{url}             
\usepackage{amssymb}

\newcommand{\Rmnum}[1]{\expandafter\@slowromancap\romannumeral #1@}

\usepackage{lineno,xcolor}
\begin{document}
\begin{article}
\begin{opening}
\title{Global Solar Free Magnetic  Energy  and Electric Current Density Distribution of  Carrington Rotation 2124}
\author{Tilaye~\surname{Tadesse}$^{1}$,
         T.~\surname{Wiegelmann}$^{2}$,
          Alexei A.~\surname{Pevtsov}$^{3}$,
         P. J. ~\surname{MacNeice}$^{1}$,      
         S. ~\surname{Gosain}$^{3}$
        }
\runningauthor{T.~Tadesse et al.}
\runningtitle{Global Solar Free Magnetic  Energy and Electric Current Density  Distribution}
 \institute{$^{1}$ NASA, Goddard Space Flight Center, Code 674, Greenbelt, MD 20771, U.S.A.
                     email: \url{tilaye.tadesse.asfaw@nasa.gov },  email: \url{peter.j.macneice@nasa.gov } \\
                 $^{2}$ Max Planck Institut f\"{u}r Sonnensystemforschung, Max-Planck Str. 2, D--37191 Katlenburg-Lindau, Germany,
                      email: \url{wiegelmann@mps.mpg.de}\\
                     $^{3}$ National Solar Observatory, Sunspot, NM 88349, U.S.A.
                     email: \url{sgosain@nso.edu}, email: \url{ah=80Mm@nso.edu} \\
    }

\begin{abstract}
Solar eruptive phenomena, like flares and coronal mass ejections(CMEs) are governed by magnetic fields. To describe
the structure of these phenomena one needs information on the magnetic flux density and the electric current density vector components in three 
dimensions throughout the atmosphere. However, current spectro-polarimetric measurements  typically limit the determination 
of the vector magnetic field only to the photosphere. Therefore, there is considerable interest in accurate modeling of the 
solar coronal magnetic field using photospheric vector magnetograms as boundary data. In this work, we model the coronal magnetic 
field for global solar atmosphere using a nonlinear force-free field(NLFFF) extrapolation codes implemented to a synoptic maps of photospheric 
vector magnetic field synthesized from {\it Vector Spectromagnetograph (VSM)} on {\it Synoptic Optical Long-term Investigations of the Sun (SOLIS)} 
as boundary condition. Using the resulting three dimensional magnetic field, we calculate the three dimensional electric current density 
and free magnetic energy throughout the solar atmosphere for Carrignton rotation 2124. We found that spatially,
the low-lying, current-carrying core field demonstrates strong concentration of free energy in the AR core, from the photosphere 
to the lower corona; the coronal field becomes slightly more sheared in the lowest layer and it relaxes to the potential field configuration with height. 
The free energy density appears largely co-spatial with the electric current distribution. 
\end{abstract}
\end{opening}

\section{Introduction}
Extreme solar activity is powered by magnetic energy \cite{Forbes:2000}. Therefore, our understanding of the energy storage and release 
mechanism of solar eruptive events is strongly dependent on knowledge of their magnetic field configurations \cite{valori:2005,Wiegelmann:2012W}. 
Flux emergence and shearing motion may introduce strong electric currents and inject energy to the active region (AR) corona. Nonpotential 
magnetic field topology carries electric current and free magnetic energy. Free magnetic energy is one of the main indicators used in 
space weather forecast to predict the eruptivity of active regions, and it can be converted into kinetic and thermal energies  as the result of eruption\cite{Regnier:2007}. 
Therefore, the three-dimensional (3D) structure of magnetic fields and electric currents in the pre-eruptive corona volume, are requirements 
to study those eruptive solar events. Thus the detailed aspects of the coronal field like twist, shear, loop structure, and connectivity have 
to be known to a high degree of accuracy. Unfortunately, due to the extremely low density and high temperature of the corona, measurements of 
the magnetic field are restricted to lower layers of the solar atmosphere. Although measurements of magnetic fields in the chromosphere and 
the corona have been considerably improved in recent decades \cite{Lin:2000,Liu:2008}, further developments are needed before accurate data are 
routinely available. Another problem with these methods is that they do not provide the field in 3D and the measurements have a line-of-sight
integrated character \cite{Judge:2013}. As an alternative to direct measurement of the three dimensional coronal magnetic field, numerical modeling are used to 
infer the field strength in the higher layers of the solar atmosphere from the measured photospheric magnetic field. 

Due to the low value of the ratio of gas pressure to magnetic pressure (plasma $\beta$ ), the solar corona is magnetically dominated and the 
field is not affected by the plasma that it carries. The coronal field is often said to beâ force-free, i.e., free of forces other than the 
balancing electromagnetic forces that it exerts on itself \cite{Gary:2001}. Force-free condition assumes that the corona is static and free of 
Lorentz forces. This means that any electric currents must be aligned with the magnetic field. Force-free coronal magnetic fields are defined 
entirely by requiring that the field has no Lorentz force and is divergence free (the solenoidal condition):
\begin{equation}
   (\nabla \times\textbf{B})\times\textbf{B}=0 \label{one}
\end{equation}
\begin{equation}
    \nabla \cdot\textbf{B}=0 \label{two}
 \end{equation}
where $\textbf{B}$ is the magnetic field. Equation~(\ref{one}) states that the Lorentz force vanishes (as a consequence of $\textbf{J}\parallel \textbf{B}$, 
where $\textbf{J}$ is the electric current density) and Equation~(\ref{two}) describes the absence of magnetic monopoles. Using Equations~(\ref{one}) and 
(\ref{two}) as constraints equations, one can calculate the magnetic field density in a corona volume for photospheric measurements. Nonlinear 
force-free magnetic field (NLFFF) extrapolation techniques are the best mathemathical modeling tools currently being used to estimate the magnetic 
field in the higher solar atmosphere \cite{Amari:2010,Aschwanden:2012,Contopoulos:2013,He:2008,Malanushenko:2012,Jiang:2012,Song:2007,Tadesse:2009,Tadesse:2013,Wheatland:2000,Wheatland:2009,Wiegelmann:2004,Wiegelmann:2007}.  For more detailed explanations of NLFFF methods, we direct the reader to the recent review by \inlinecite{Wiegelmann:2012W}.

\begin{figure*}
\includegraphics[viewport=0 10 510 710,clip,height=21.5cm,width=17.5cm]{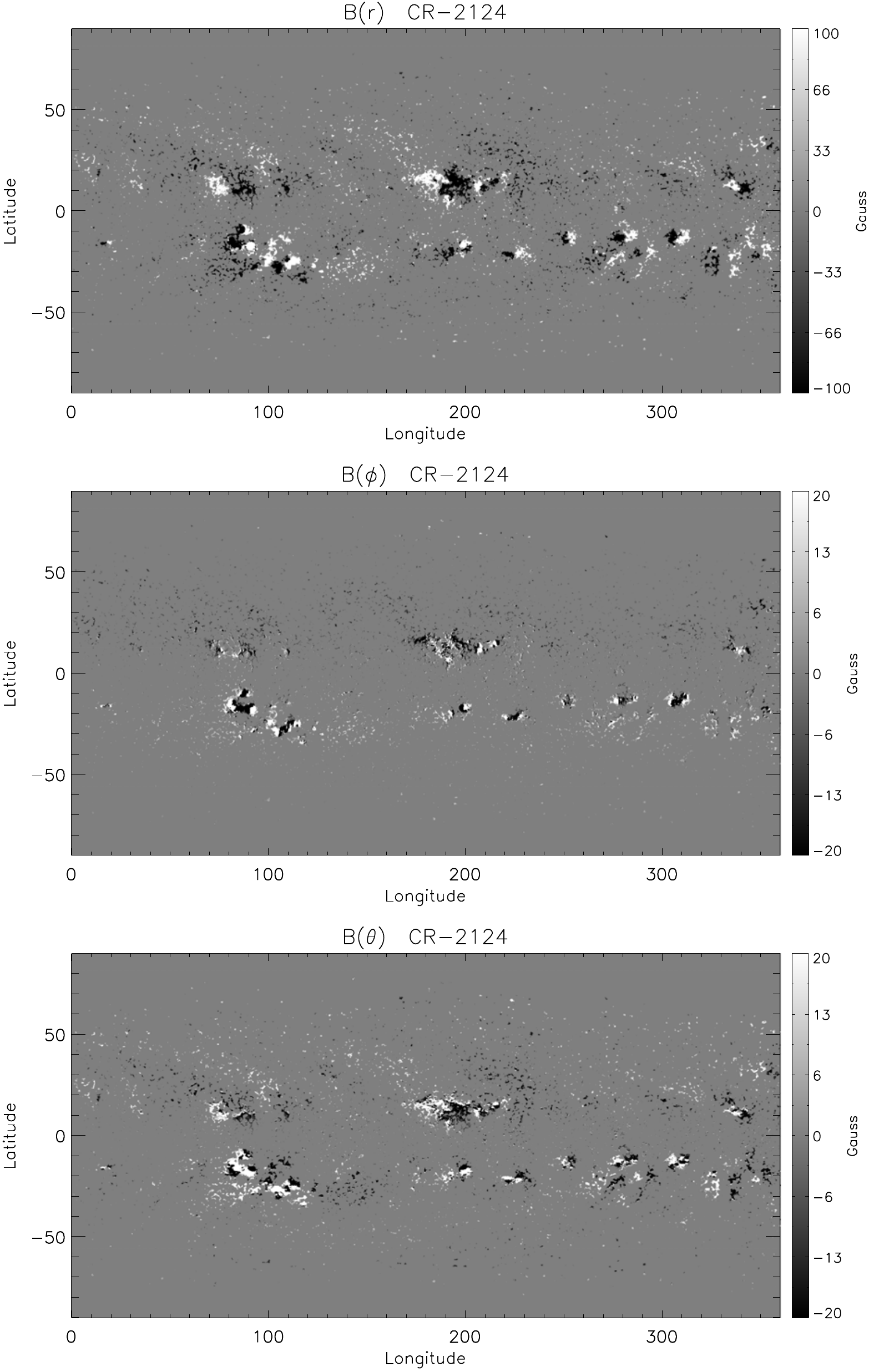}
\caption{Synoptic Carrington maps of the vector magnetic field components are shown for CR-2124. The panels from top to 
bottom show the distribution of the $B(r)$, $B(\phi)$ and $B(\theta)$ components, respectively. The $B_{r}$ map is scaled 
between $\pm$ 100 G, and the $B_{\phi}$ and $B_{\theta}$ maps are scaled to $\pm$ 20 G. The positive values of $B_{r}$, 
$B_{\phi}$ and $B_{\theta}$ point, respectively, upward, to the right (westward) and southward.}
\label{fig1}
\end{figure*}

Electric currents are thought to play an important role for the energetics of the chromosphere and the corona of the Sun. Twisted magnetic 
field line implies the presence of field-aligned electric currents (but see \inlinecite{h=80Mm:1997} for counter example of potential field lines that appear highly sheared). 
The existence of electric currents  implies departure of the magnetic structure from its minimum-energy, current-free (potential) configuration. Therefore, free magnetic 
energy is associated with electric currents flowing in the solar corona. 
Having the three-dimensional structure of the vector magnetic field,  one can compute the vector current density as the curl of the field.  On the largest scales, 
the corona of a whole hemisphere was found to be magnetically coupled \cite{Schrijver:2009,h=80Mm:2001}.
Therefore, it appears prudent to implement a NLFFF procedure in spherical geometry  \cite{Amari:2013,Tadesse:2013ad} for use when global 
scale boundary data are available (see Figure~\ref{fig1}), such as synoptic maps of photospheric vector magnetic field synthesized from {\it Vector Spectromagnetograph} 
(VSM) on {\it Synoptic Optical Long-term Investigations of the Sun} (SOLIS) \cite{Gosain:2013}. In this work, we apply spherical NLFFF 
procedure to synoptic vector map of Carrington rotation 2124 to reconstruct 3D magnetic field solutions (potential and NLFFF) from which we infer the distributions of 
 free magnetic energy and electric current density in corona volume globally.
\section{Optimization Procedure for Global Nonlinear Force-Free Field Model}
\inlinecite{Wheatland:2000} derived an optimization approach based on a pseudo-energy integral which is minimized by a divergence-free, 
force-free field [Equations (\ref{one}) and (\ref{two})] using Cartesian geometry. The variation of the integral with respect to the variation 
of the magnetic field gives a pseudo-force, which can be used to dynamically drive the system from a potential field toward a force-free field that 
minimizes the pseudo-energy integral, subject to the required boundary conditions at the photospheric level. The optimization method 
requires that the magnetic field is given on all (6 for a rectangular computational box) boundaries. This causes a serious limitation of the
method because such data are only available for model configurations. Later on, \inlinecite{Wiegelmann:2004} improved the method in a such 
away that it can reconstruct the magnetic field only from photospheric vector magnetograms by diminishing the effect of the top and lateral 
boundaries on the magnetic field inside the computational box.

Since extrapolation codes of Cartesian geometry for medelling the magnetic field in the corona do not take the curvature of the Sun's surface into 
account, \inlinecite{Wiegelmann:2007} extended the method into spherical coordinates in a such way that it can handle the the Sun, globally. 
Later on, \inlinecite{Tadesse:2009} developed code for the extrapolation of nonlinear force-free coronal magnetic fields in spherical coordinates 
over a restricted area of the Sun \cite{Tadesse:2012,Tadesse:2012a,Tadesse:2013sol}. In this work, we use the optimization procedure for functional 
$L$ in spherical geometry for the whole Sun globaly \cite{Wiegelmann:2007,Tadesse:2009}, which enforces a minimal deviation between the photospheric 
boundary of the model field $\textbf{B}$ and the measured field value $\textbf{B}_{\rm obs}$ by adding an appropriate 
surface integral term $L_{\rm photo}$ \cite{Wiegelmann:2010,Tadesse:2011} as in the following equation:
\begin{equation}L=L_{\rm f}+L_{\rm d}+\nu L_{\rm photo} \label{4}
\end{equation}
\begin{displaymath} L_{\rm f}=\int_{V}B^{-2}\big|(\nabla\times {\textbf{B}})\times
{\textbf{B}}\big|^2  r^2\sin\theta dr d\theta d\phi
\end{displaymath}
\begin{displaymath}L_{\rm d}=\int_{V}\big|\nabla\cdot {\textbf{B}}\big|^2
  r^2\sin\theta dr d\theta d\phi
\end{displaymath}
\begin{displaymath}L_{\rm photo}=\int_{S}\big(\textbf{B}-\textbf{B}_{\rm obs}\big)\cdot\textbf{W}(\theta,\phi)\cdot\big(
\textbf{B}-\textbf{B}_{\rm obs}\big) r^{2}\sin\theta d\theta d\phi
\end{displaymath}
where $L_{\rm f}$ and $L_{\rm d}$ measure how well the force-free equation [Equation~(\ref{one})] and divergence-free condition [Equation~(\ref{two})] 
are fulfilled, respectively. We implement the surface integral term, $L_{\rm photo}$, in Equation~(\ref{4}) to work with boundary 
data of different noise levels and qualities \cite{Wiegelmann:2010,Tadesse:2011}.  This  allows deviations between the model field $\textbf{B}$ and the 
input field, {\it i.e.} the observed $\textbf{B}_{\rm obs}$ surface field, so that the model field can be iterated closer to a force-free solution. $\textbf{W}(\theta,\phi)$ is a 
space-dependent diagonal matrix whose elements $(w_{los},w_{trans},w_{trans})$ are inversely proportional to the estimated squared measurement error of the respective 
field components. Because the line-of-sight photospheric magnetic field is measured with much higher accuracy than the transverse field, we typically set the component 
$w_{los}$ to unity, while the transverse components of $w_{trans}$ are typically small but positive. In regions where transverse field has not been measured or where the 
signal-to-noise ratio is very poor, we set $w_{trans}=0$. In order to control the speed with which the lower boundary condition is injected during the NLFFF extrapolation, we have 
used the Lagrangian multiplier of $\nu=0.001$ as suggested by \inlinecite{Tadesse:2013}.  

We use spherical preprocessing procedure to remove most of the net force and torque from the synoptic vector data so that the boundary can be more consistent with the force-free 
assumption \cite{Wiegelmann:2006,Tadesse:2009}.  The potential field solution has been used to initialize NLFFF code. To calculate NLFFF solutions,  we minimize the functional 
given by Equation~(\ref{4}) using preprocessed photospheric  boundary data.  We use a uniform spherical grid $r$, $\theta$, $\phi$ with $n_{r}=225$, $n_{\theta}=375$, and $n_{\phi}=
425$ grid points in the direction of radius, latitude, and longitude, respectively. We aim to compute the whole sphere with the field of view of $[r_{\rm{min}}=1R_{\odot}:r_{\rm{max}}=
1.5R_{\odot}]\times[\theta_{\rm{min}}=-90^{\circ}:\theta_{\rm{max}}=90^{\circ}]\times[\phi_{\rm{min}}=0^{\circ}:\phi_{\rm{max}}=360^{\circ}]$.  To avoid the mathematical singularities 
at the poles, we do not use grid points exactly at the south and north pole, but set them half a grid point apart at $\theta_{min} = -90^{\circ}+d\theta/2$ and $\theta_{max} =
90 ^{\circ}-d\theta/2$ (see,  \opencite{Wiegelmann:2007}).  
\section{Results}
\begin{figure}[htp!]
  \includegraphics[viewport=5 5 570 565,clip,height=14.8cm,width=15.5cm]{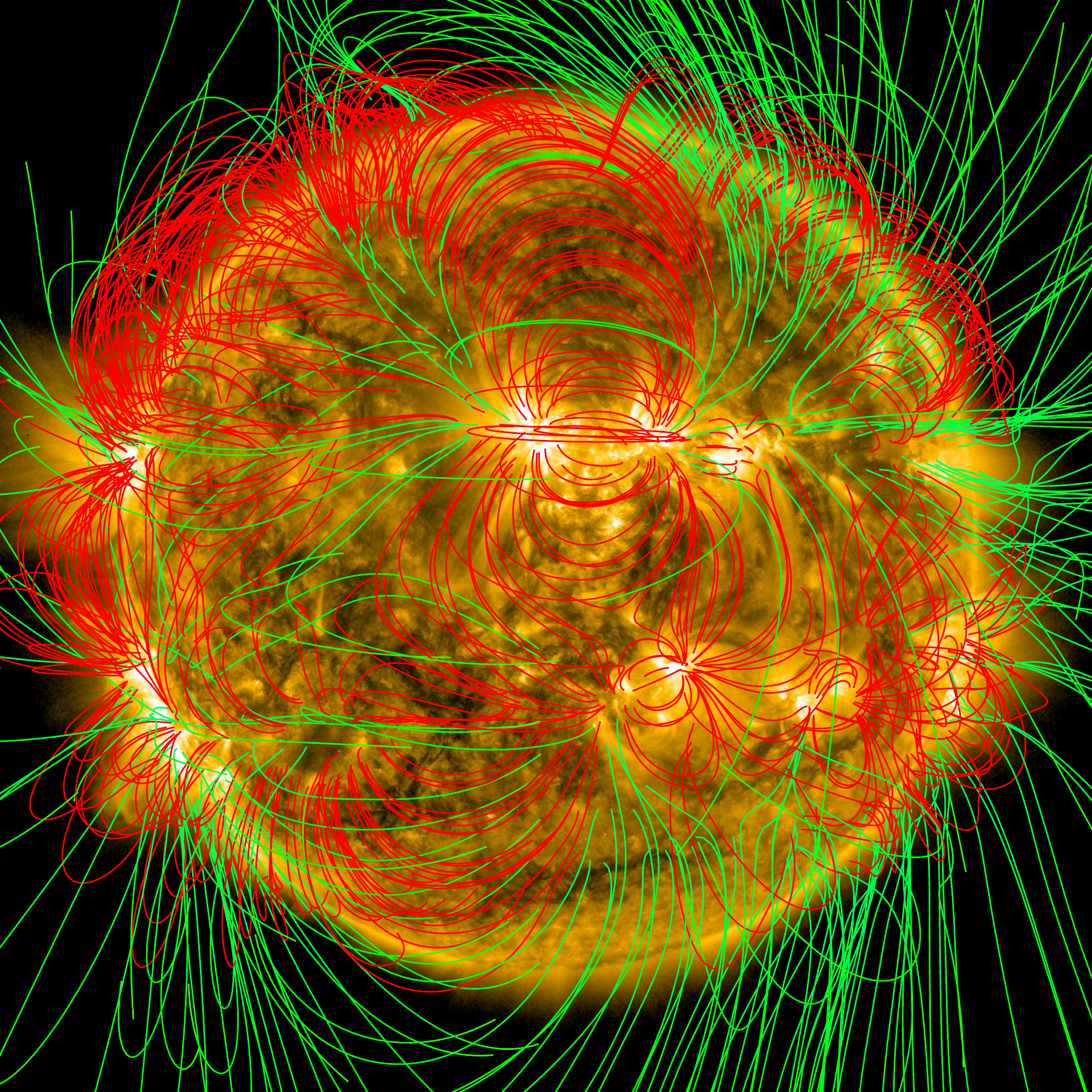}
   \caption{Global field lines of the NLFFF model overlaid on the AIA 171\AA{} image.  Green and red lines represent open and closed magnetic field lines, respectively.}\label{fig2}
 \end{figure}
\begin{figure}
     \subfloat[]{\includegraphics[ viewport=15 5 610 455,clip,height=4.0cm,width=6.0cm]{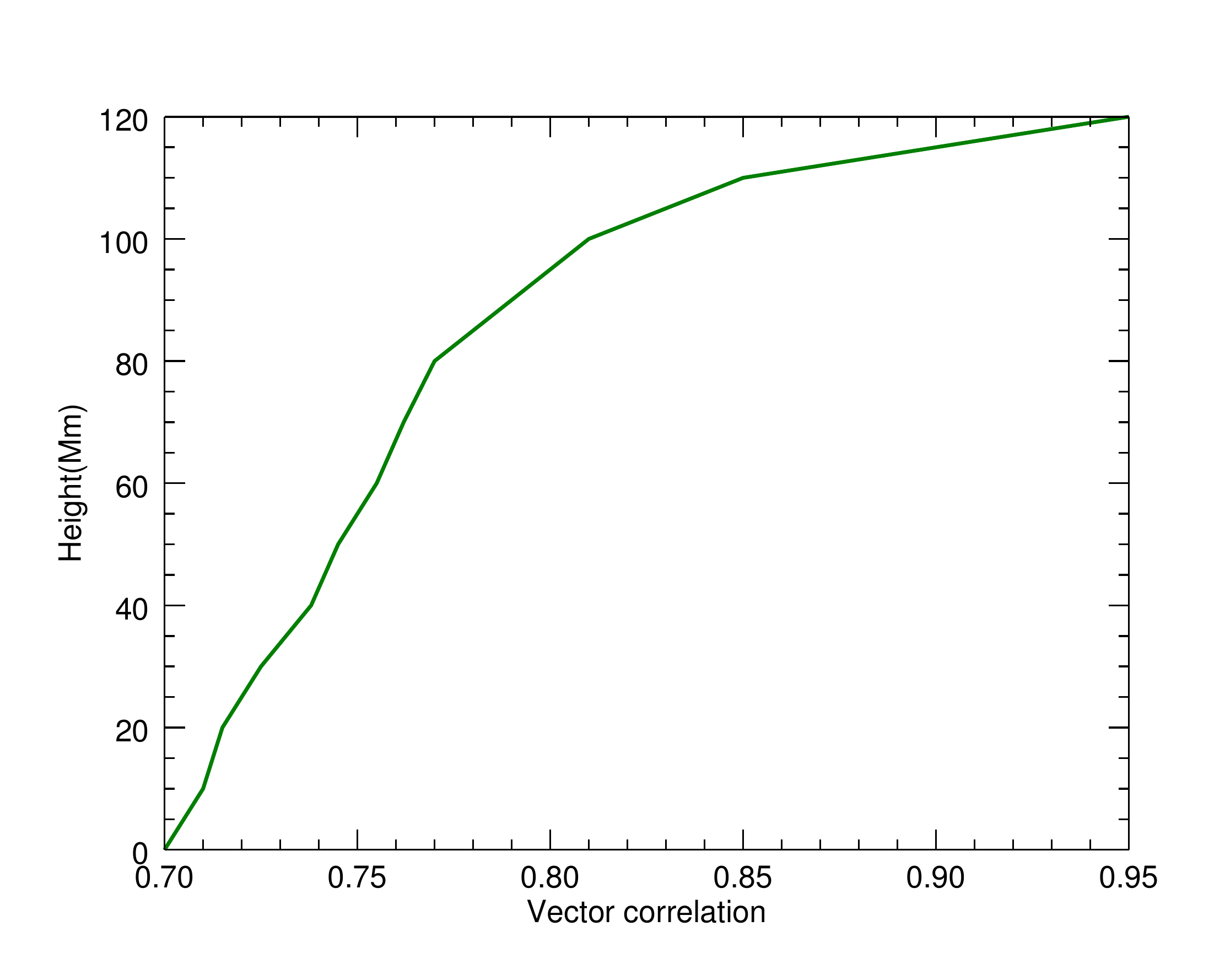}}
      \subfloat[]{\includegraphics[ viewport=15 5 610 455,clip,height=4.0cm,width=6.0cm]{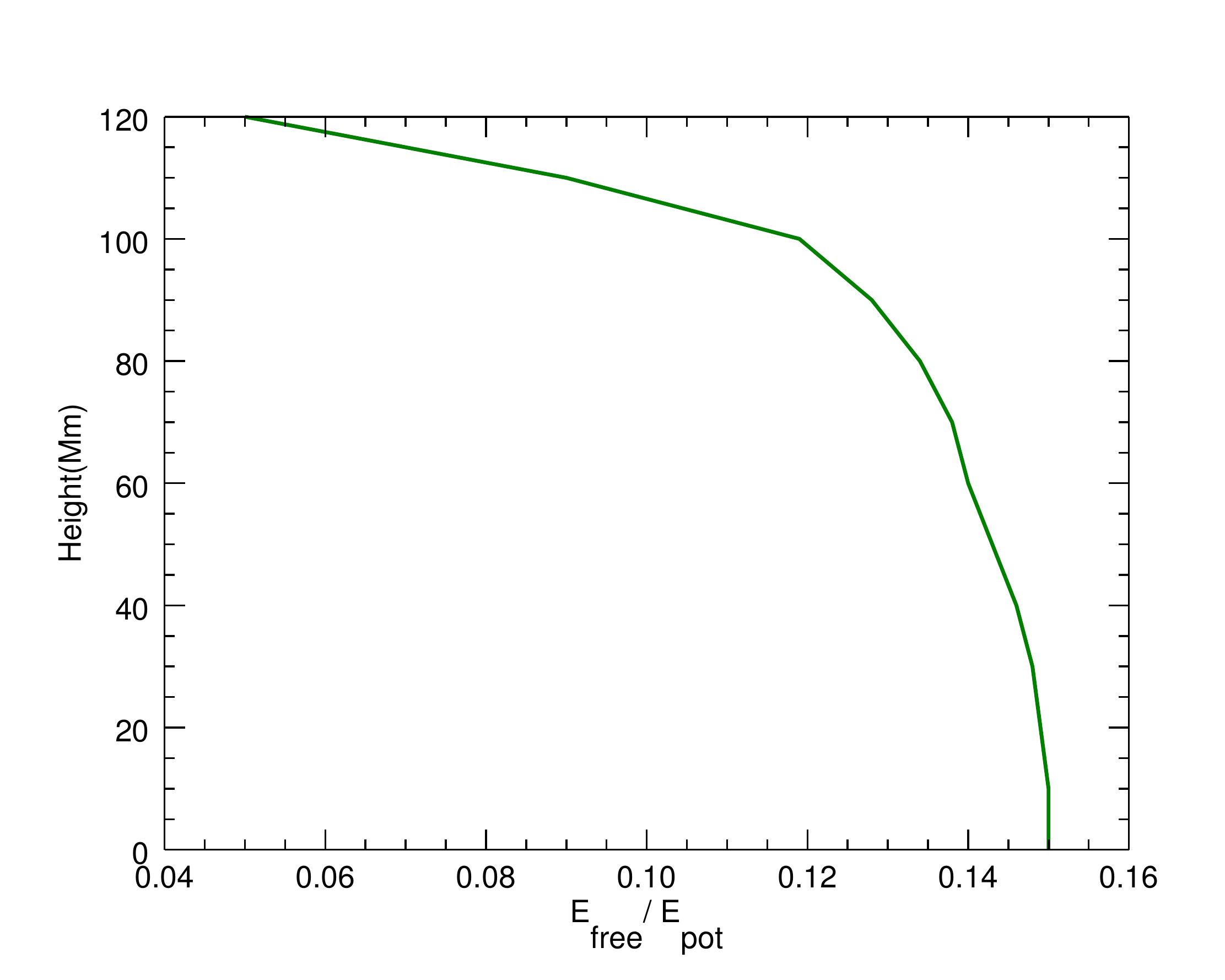}}
      \caption{(a) Plot of height variations of surface vector correlation between the potential field and NLFFF solutions. (b)  Plot of height variations of ratio of free magnetic energy to potential  magnetic energy.}\label{fig3}
 \end{figure}
\begin{figure}
\subfloat[]{\includegraphics[viewport=10 105 530 410,clip,height=8.2cm,width=14.2cm]{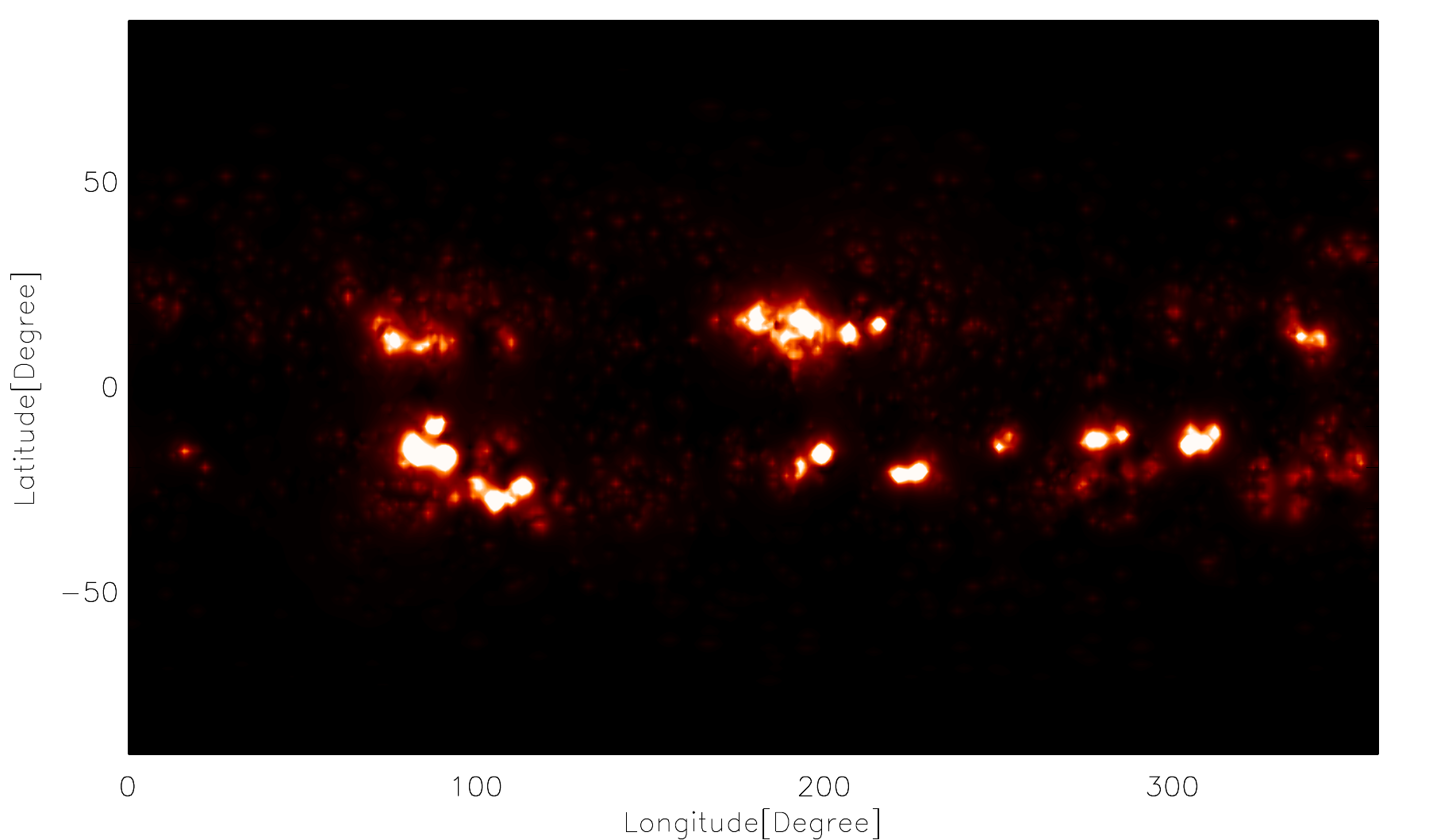}}\\
\subfloat[]{\includegraphics[viewport=10 105 530 410,clip,height=8.2cm,width=14.2cm]{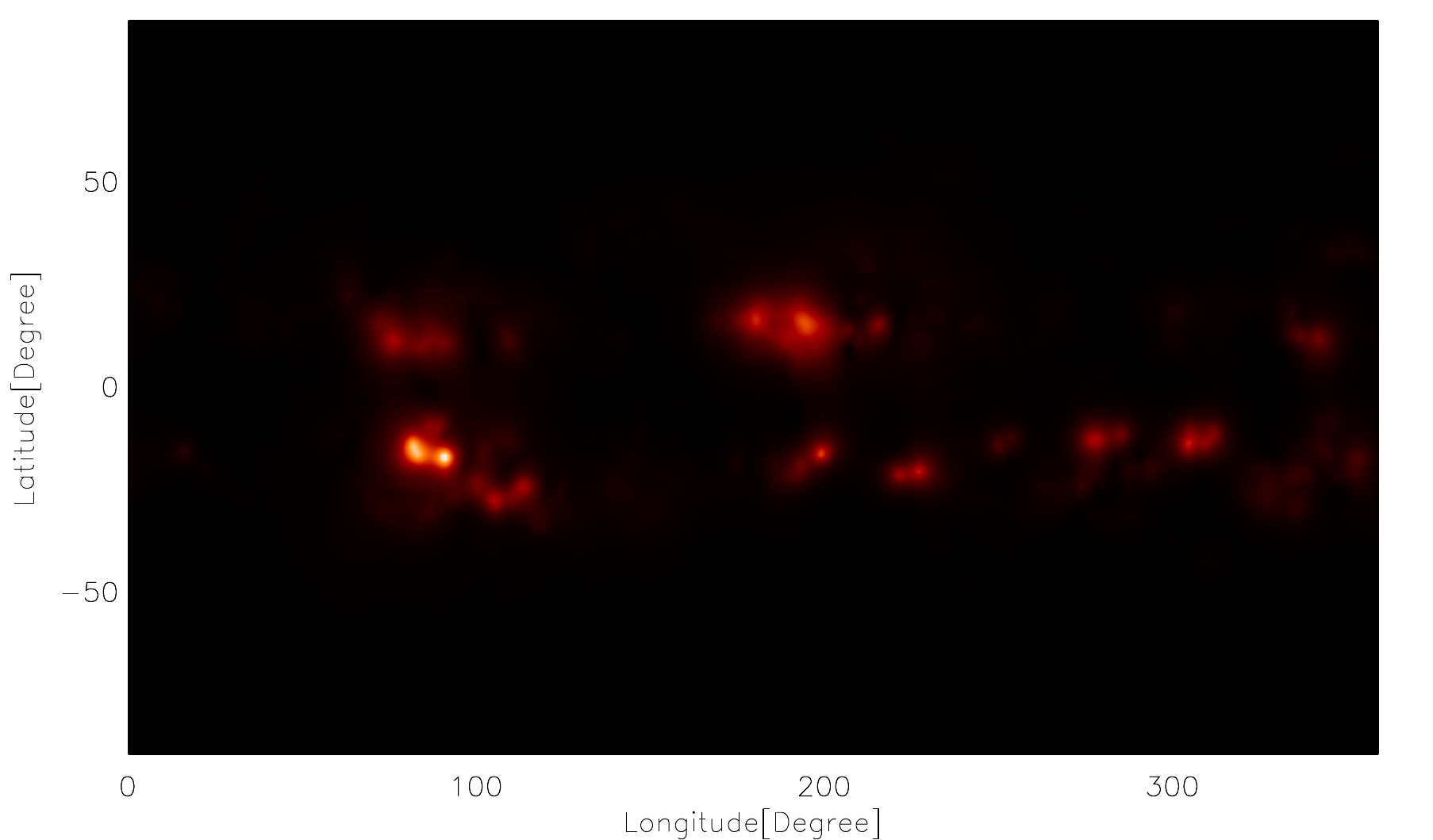}}\\
\includegraphics[viewport=10 45 530 105,clip,height=1.5cm,width=14.2cm]{free_energy2.pdf}
\caption{Synoptic maps of the global free magnetic energy relative to the potential energy at (a) $h=0$ (photosphere) and (b) at $h=80Mm$ from the photosphere.}
\label{fig4}
\end{figure}
 In this section we apply the spherical optimization procedure  to synoptic maps of photospheric vector magnetic field synthesized from {\it Vector Spectromagnetograph} (VSM) on {\it Synoptic Optical Long-term Investigations of the Sun} (SOLIS). The main purpose of this work is to study the distributions of free magnetic energy and electric current density in 
 the global corona for Carrington rotation 2124 (May 2012 -- June 2012) using NLFFF extrapolation. Two physical measures,  the free magnetic energy and  electric current density  were calculated from  the extrapolated fields to quantitatively evaluate the structure of the reconstructed 3D magnetic field for global corona. 
\begin{figure}
\includegraphics[viewport=10 35 540 310,clip,height=5.2cm,width=12.2cm]{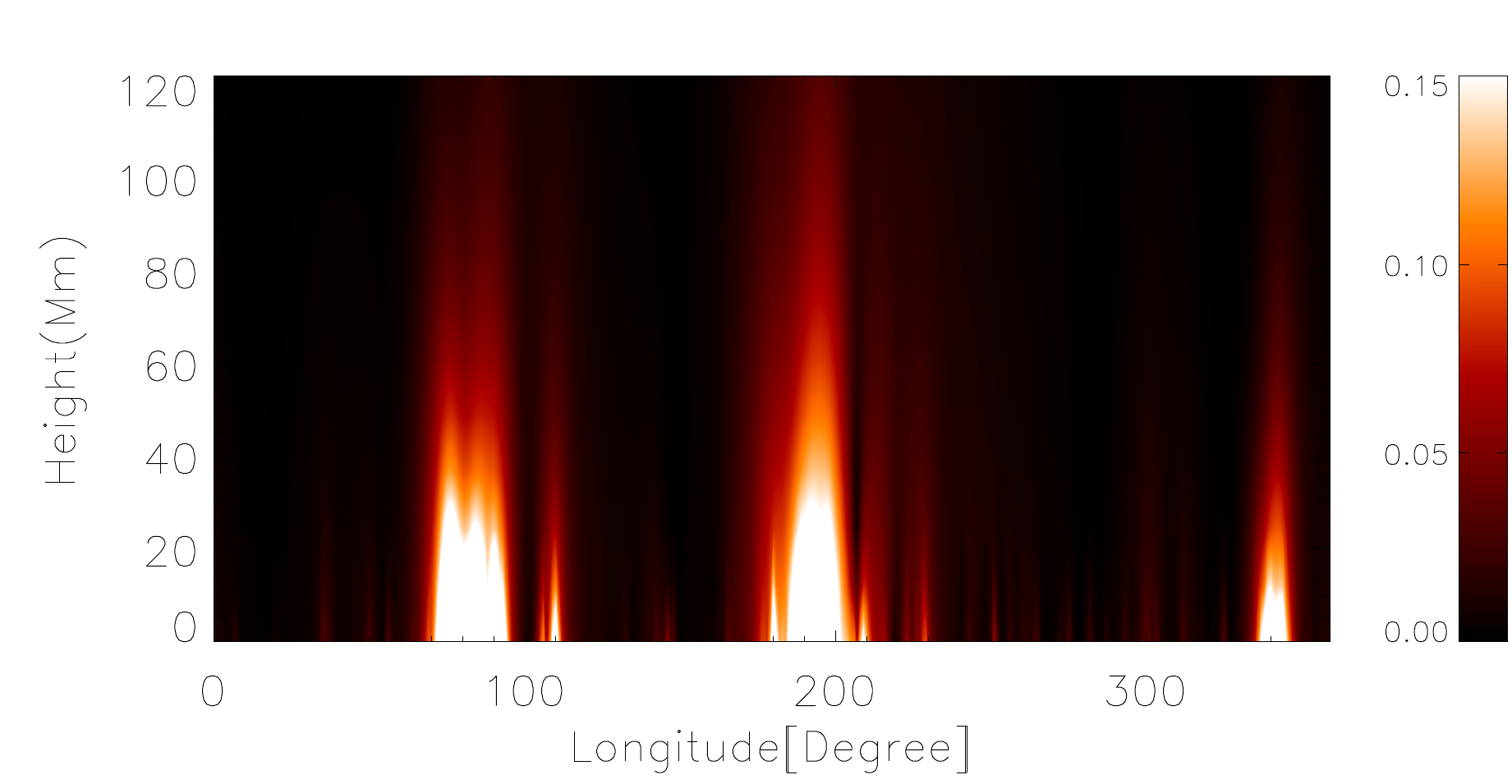}
\caption{Free magnetic energy density map on the latitudinal plane $\theta=15^{\circ}$ for global NLFFF field relative to potential field (increasing density from black to white).}
\label{fig5}
\end{figure}
\subsection{Global Free Magnetic Energy}
Once the 3D magnetic configuration is reconstructed, we first have to check if characteristic magnetic structures can be obtained in comparison with observations. 
Therefore, the field line configurations of the  extrapolated magnetic fields are compared with an AIA 171 \AA{}  coronal loop observations in Figure~\ref{fig2}. The Figure 
shows the open and closed field lines of the extrapolated magnetic fields  overlying the corresponding AIA image.  The comparison shows that the field lines of the modeling fields generally agree with the coronal loop images. We now consider the distribution of the magnetic free energy based on the NLFFF extrapolation. The potential field is a minimum of magnetic energy for a given distribution of the vertical or radial magnetic field component on the photosphere \cite{Aly:1984}. Therefore there is no free magnetic energy, no shear 
and/or twisted field lines in a potential field configuration. For these reasons, the nonlinear force-free field is more adequate to describe better the nature of the corona as it contains 
free magnetic energy and sheared and twisted flux bundles.  We estimate the free magnetic energy, the difference between the extrapolated NLFFF and the potential field with the 
same normal boundary conditions in the photosphere \cite{Regnier:2007,Regnier:2007b}. The free magnetic energy is a measure of the magnetic energy that can be stored in the magnetic configuration or released during an eruptive or reconnection event. We therefore estimate the upper limit to the free magnetic energy associated with coronal currents of 
the form
\begin{equation}
E_\mathrm{free}=\frac{1}{8\pi}\int_{V}\Big(B_{\rm nlff}^{2}-B_{\rm pot}^{2}\Big)r^{2}\sin\theta dr d\theta d\phi, \label{ten}
\end{equation}
where $B_{\rm nlff}$ and $B_{\rm pot}$ are NLFFF and potential field solutions, respectively.  Our result for the estimation of free-magnetic energy shows that the NLFFF 
model has $15\%$ more energy than the corresponding potential field model calculated from the same radial field boundary. 

In order to quantify the degree of deviations between the potential and NLFFF vector field solutions in the global corona volume, we use the 
vector correlation metric ($C_{\rm vec}$) which is also used analogous to the standard correlation coefficient for scalar functions. The correlation was calculated 
\cite{Schrijver:2006} from
\begin{equation}
C_\mathrm{ vec}= \frac{ \sum_i \textbf{v}_{i} \cdot \textbf{u}_{i}}{ \Big( \sum_i |\textbf{v}_{i}|^2 \sum_i
|\textbf{u}_{i}|^2 \Big)^{1/2}}, \label{6}
\end{equation}
where $\textbf{v}_{i}$ and $\textbf{u}_{i}$ are the vectors at each grid point $i$. If the vector fields are identical, then $C_{\rm vec}=1$; 
if $\textbf{v}_{i}\perp \textbf{u}_{i}$ , then $C_{\rm vec}=0$.  We calculated vector correlations of the potential and NLFFF solutions at different layers in corona volume and found that 
the correlation increases with height (see, Figure~\ref{fig3}a).  The result shows that the coronal field loop becomes slightly more sheared in the lowest layer, relaxes to the potential field configuration with height. Similarly we calculated the ratio of total surface free magnetic  to potential field energy over different layers in corona volume globally and found that 
the ratio decreases with height as shown in Figure~\ref{fig3}b.  For illustration, we plot total surface free magnetic energy relative to the potential one at two layers. Figures~\ref{fig4}(a) 
and (b) show the contour plots of the global free magnetic energy relative to the potential energy on the photosphere and  on the layer at height $h=80Mm$ above the photosphere, respectively.  In Figure~\ref{fig5}  we plot the free magnetic energy density  on the latitudinal plane $\theta=15^{\circ}$ for global NLFFF field relative to potential field and we found that spatially, the low-lying, current-carrying core field demonstrates strong concentration of free energy in the AR core, from the photosphere to the lower corona. 
\begin{figure}
\begin{center}
\subfloat[]{\includegraphics[viewport=10 95 530 415,clip,height=8.2cm,width=14.2cm]{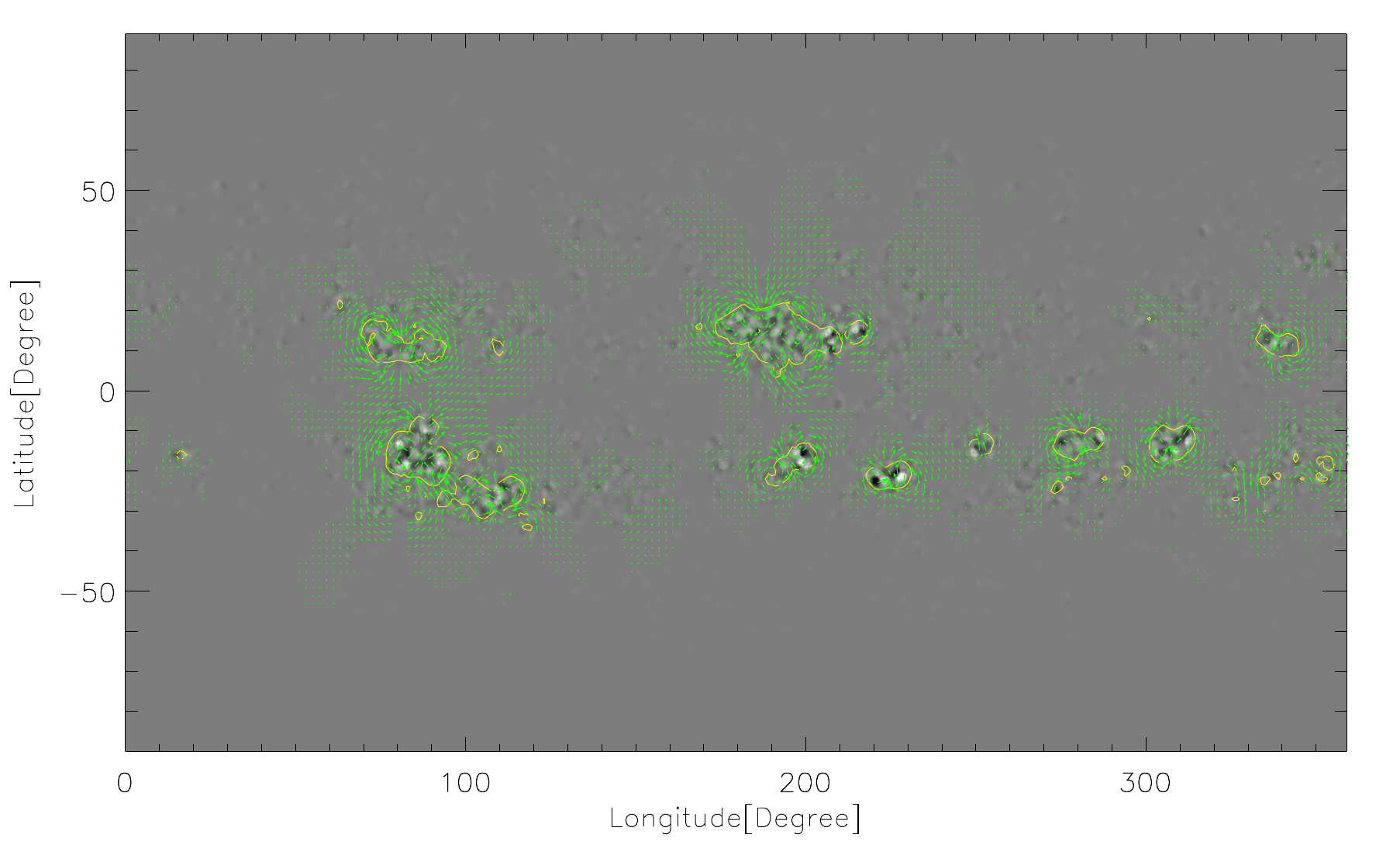}}\\
\subfloat[]{\includegraphics[viewport=10 95 530 415,clip,height=8.2cm,width=14.2cm]{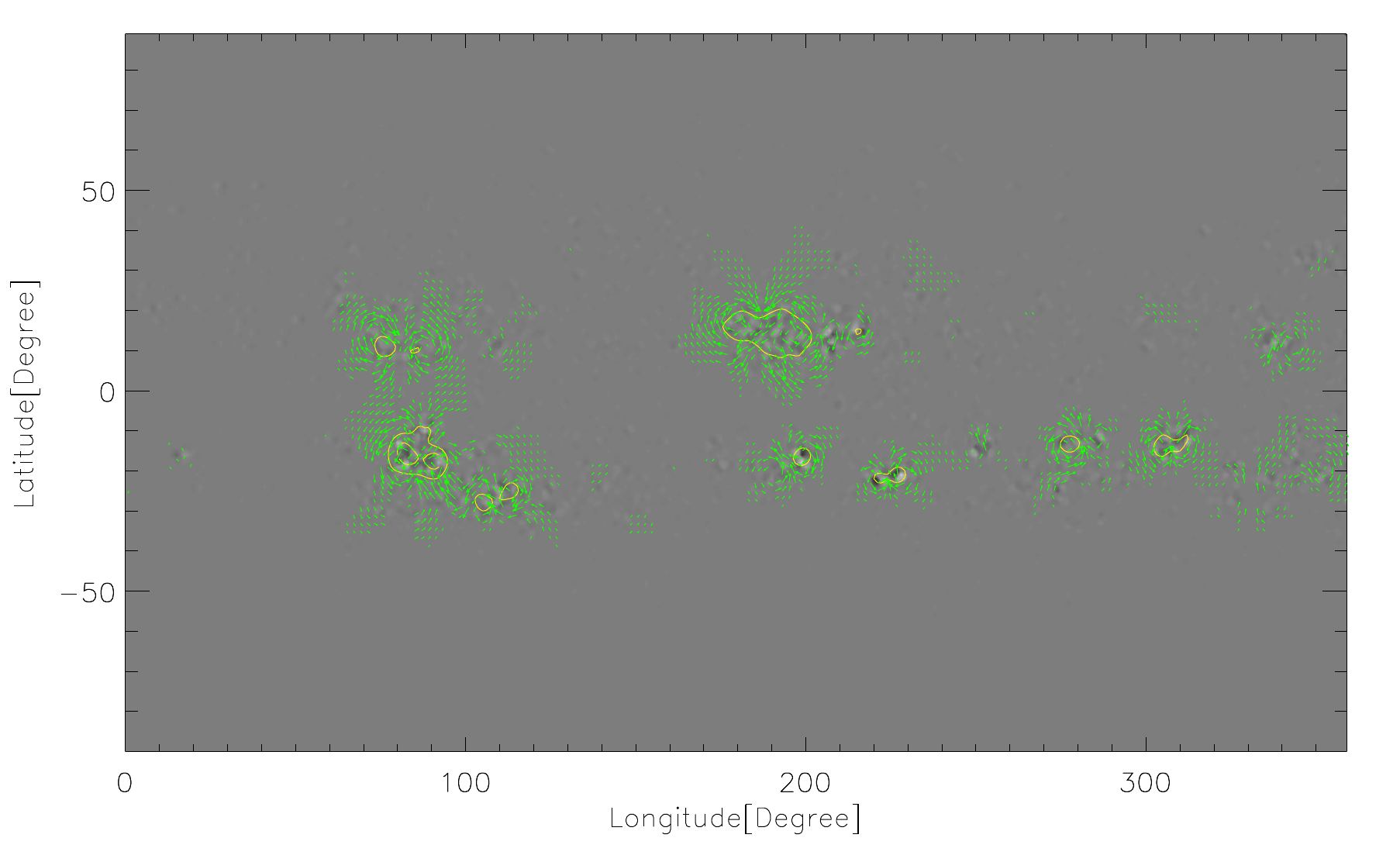}}\\
\includegraphics[viewport=10 35 530 95,clip,height=1.5cm,width=14.2cm]{current_free_energy_chromo.pdf}
\caption{Synoptic maps of  electric current density distributions at (a) $h=0$ (photosphere) and (b) $h=80Mm$ from the photosphere. The color coding  and the 
green arrows show $J_{r}$ and transverse components of electric current densities, respectively. The contour lines indicate the locations of free magnetic energy concentration.}
\label{fig6}
\end{center}
\end{figure}
\begin{figure}
\begin{center}
\includegraphics[viewport=30 90 760 470,clip,height=6.5cm,width=13.2cm]{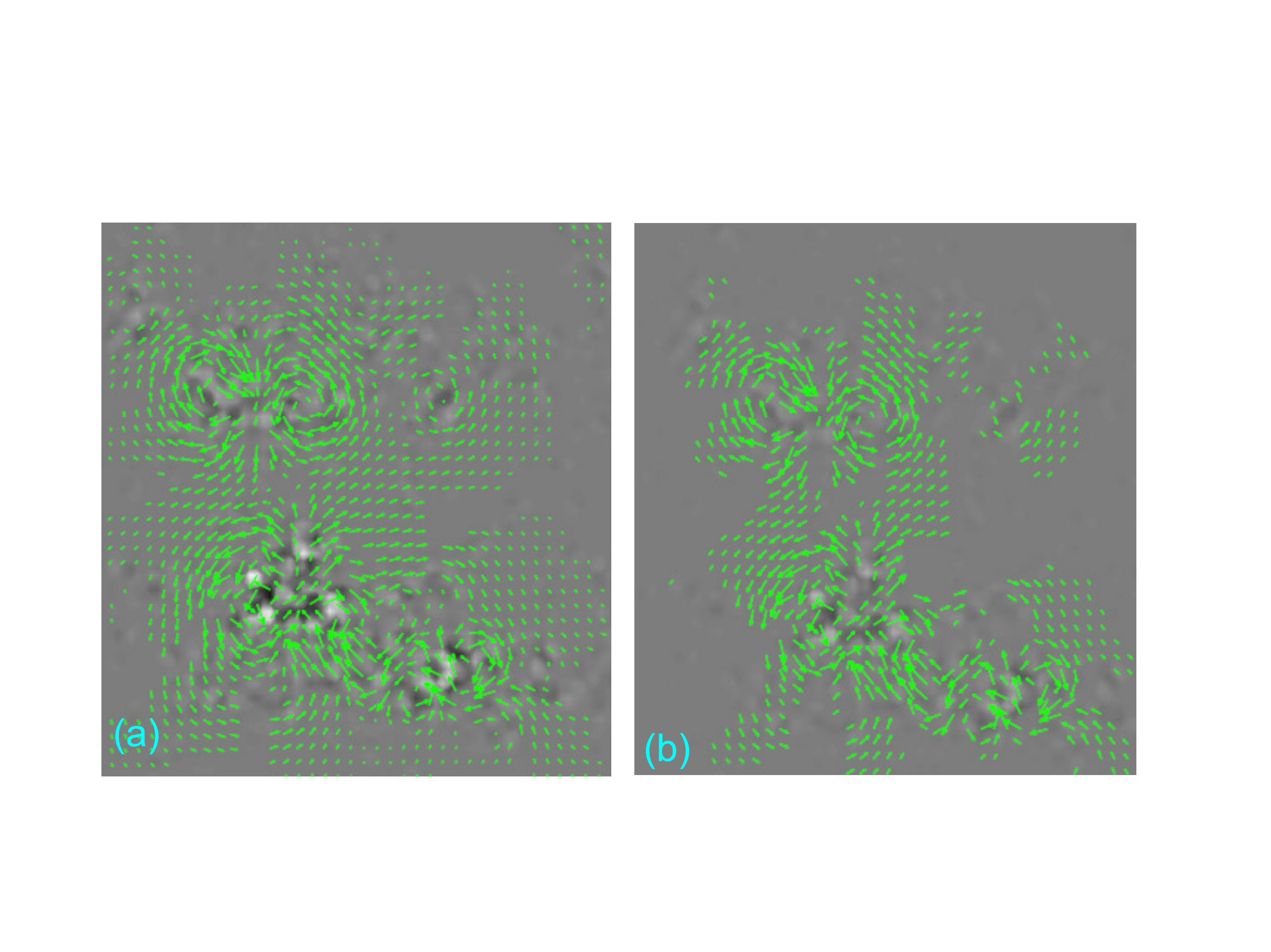}\\
\includegraphics[viewport=30 35 530 95,clip,height=1.5cm,width=13.2cm]{current_free_energy_photo.pdf}
\caption{Surface electric current density distributions at (a) $h=0$ (photosphere) and (b) $h=80Mm$ from the photosphere. The color coding  and the green arrows show $J_{r}$ and 
transverse components of electric current densities, respectively. }
\label{fig7}
\end{center}
\end{figure}
\begin{figure}
\begin{center}
\includegraphics[viewport=10 10 895 470,clip,height=6.5cm,width=12.8cm]{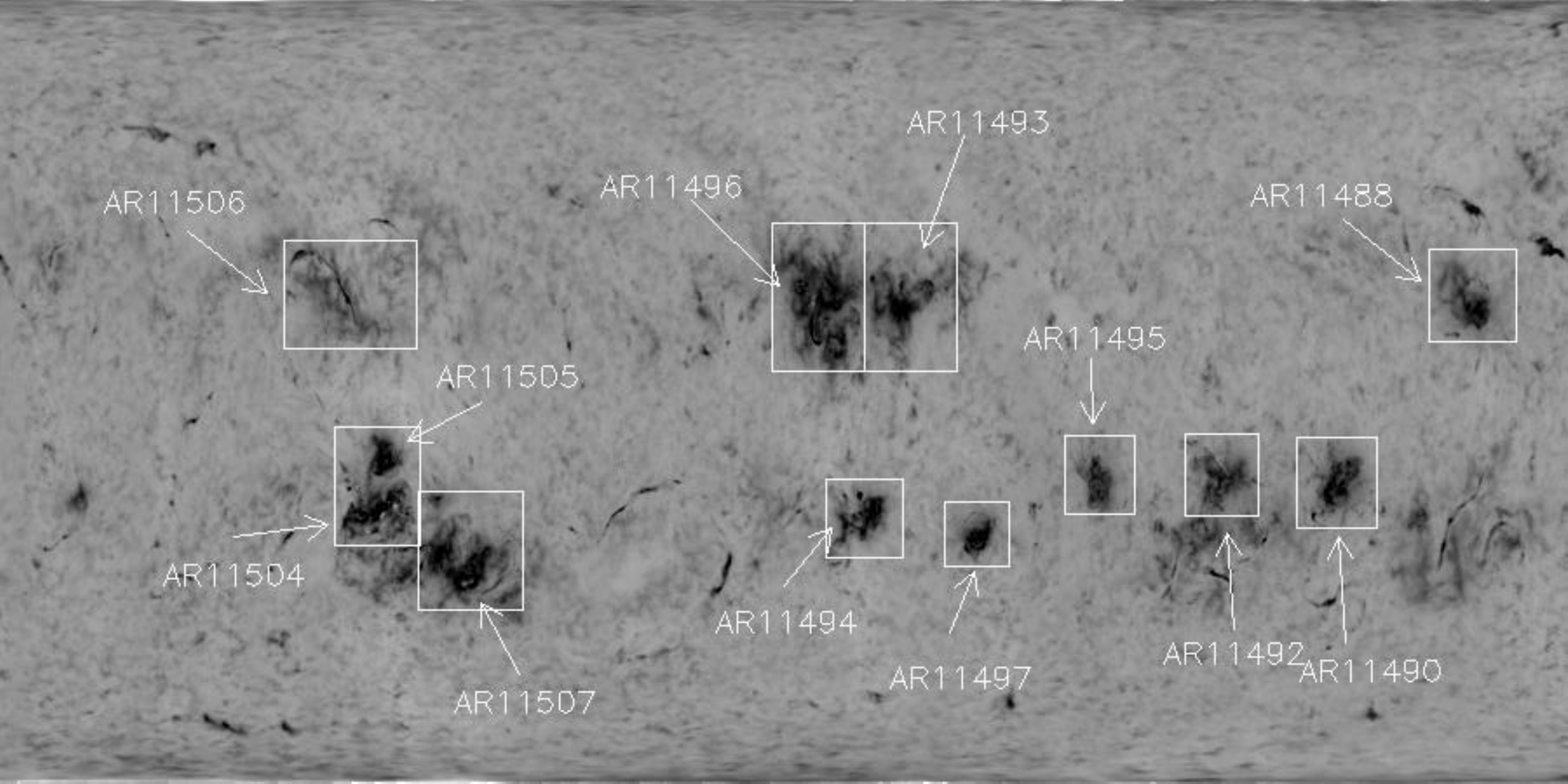}
\caption{He10830 synoptic map for Carrington Rotation 2124.}
\label{fig8}
\end{center}
\end{figure}

\subsection{Electric Current Density Distribution}
Electric current systems are key component in understanding solar activity.  Free magnetic  energy is thought to derive from the non-potential component of the magnetic field, 
which is the part associated with electric currents flowing in the solar corona. Thus, free magnetic energy is directly connected to the magnitude of the electric currents in the solar atmosphere.  Having the three-dimensional structure of the vector magnetic field, one can compute the vector current density $\textbf{J}$ as the
curl of the field: 
\begin{equation}
\textbf{J}=\frac{1}{4\pi}\nabla\times\textbf{B}, \label{7}
\end{equation}
where $\textbf{B}$ is  nonlinear force-free  3D magnetic field solution. We use finite difference method to solve for the electric current density from curl of $B$. We plot electric current density vector $\textbf{J}$ at two different  heights in Figure~\ref{fig6}. Figures~\ref{fig6}(a) and (b) show the contour plots of the global free magnetic energy and electric current densities on the photosphere and  on the layer at height $h=80Mm$ above the photosphere, respectively. Figure~\ref{fig6} shows that free energy density appears largely co-spatial with the current distribution. There is high free energy concentration above polarity inversion lines (PILs). In Figure~\ref{fig7}  shows zooming of the electric current densities of group of active regions located on the left side of Figures~~\ref{fig6}a). and b). It shows that the radial and transverse components of the current densities decrease at $h=80Mm$ height from the photosphere.
\section{Discussion }
In this study, we have investigated the distributions  free magnetic energy and electric current density associated with nonlinear force-free global magnetic field in the  corona by 
analyzing synoptic maps of photospheric vector magnetic field synthesized from Vector Spectromagnetograph (VSM) on Synoptic Optical Long-term  Investigations of the Sun (SOLIS).  The Carrington rotation number for this observation is 2124, which has been observed during 26th March - 21st June, 2012. We have used spherical NLFFF to compute the magnetic field solutions over global corona. 

We have compared the NLFFF magnetic field solutions with  the observed corresponding EUV loops.  The comparison shows that the field lines of the modeling fields generally agree with the coronal loop images. For this particular Carrington rotation we found that the free magnetic energy density is $15\%$ of the potential one and spatially,
the low-lying, current-carrying core field demonstrates strong concentration of free energy in the AR core, from the photosphere to the lower corona.  One can compare the distribution of non-potential energy with coronal brightness integrated over selected active regions. In this comparison, we use synoptic map of equivalent width of spectral line He 1083.0 nm (low solar corona) from SOLIS/VSM (Figure~\ref{fig8}).  Visually, active regions with the largest  density of free energy (per unit area) at 80Mm are AR11504, AR11496, and AR11494 (see, brightest pixels in Figure~\ref{fig4}b). These active regions also exhibit the strongest equivalent width in Figure~\ref{fig8}. This suggests that active regions with largest amount of non-potential energy are brighter in the corona.  In this study, the free energy density appears largely co-spatial with the current density distribution and  the coronal field loop becomes slightly more sheared in the lowest layer, relaxes to the potential field configuration with height.  By comparing Figures~\ref{fig7}a and b one can see that the patterns of electric currents are similar in the photosphere and 80Mm above it. These is no indication that the currents spread over larger area higher in atmosphere, while the amplitude of $J$-total clearly decreases with height. This decrease in amplitude of electric currents is in agreement with our conclusion that the magnetic field became more potential with height. It also indicates that the electric currents dissipate strongly with atmospheric height, and that there is no continuity of J through solar atmosphere. Photospheric electric currents are strongest in AR11504 and AR11497 (see Figure~\ref{fig6}a). On the other hand, the coronal brightness of AR11497 is not as high as in AR11504. This suggests only a weak correlation between the photospheric electric currents and the coronal brightness in agreement with \inlinecite{Fisher:1998}.

\section*{Acknowledgements} The authors thank the anonymous referee for helpful and detailed comments. 
Data are courtesy of NASA/SDO and the AIA and HMI science teams. This work utilizes SOLIS data obtained by the NSO Integrated Synoptic 
Program (NISP), managed by the National Solar Observatory, which is operated by the Association of Universities for Research in Astronomy (AURA), 
Inc. under a cooperative agreement with the National Science Foundation. This research was partly supported by NASA grant NNX07AU64G and by an 
appointment to the NASA Postdoctoral Program at the Goddard Space Flight Center (GSFC), administered by Oak Ridge Associated Universities through 
a contract with NASA.

\end{article}

\begin{thebibliography}{37}
\ifx \bisbn   \undefined \def \bisbn  #1{ISBN #1}\fi
\ifx \binits  \undefined \def \binits#1{#1}\fi
\ifx \bauthor  \undefined \def \bauthor#1{#1}\fi
\ifx \batitle  \undefined \def \batitle#1{#1}\fi
\ifx \bjtitle  \undefined \def \bjtitle#1{\textit{#1}}\fi
\ifx \bvolume  \undefined \def \bvolume#1{\textbf{#1}}\fi
\ifx \byear  \undefined \def \byear#1{#1}\fi
\ifx \bissue  \undefined \def \bissue#1{#1}\fi
\ifx \bfpage  \undefined \def \bfpage#1{#1}\fi
\ifx \blpage  \undefined \def \blpage #1{#1}\fi
\ifx \burl  \undefined \def \burl#1{\textsf{#1}}\fi
\ifx \href  \undefined \def \href#1#2{\textsf{#2}}\fi
\ifx \doiurl  \undefined \def
  \doiurl#1{\href{http://dx.doi.org/#1}{\textsf{#1}}}\fi
\ifx \betal  \undefined \def \betal{\textit{et al.}}\fi
\ifx \binstitute  \undefined \def \binstitute#1{#1}\fi
\ifx \bctitle  \undefined \def \bctitle#1{#1}\fi
\ifx \beditor  \undefined \def \beditor#1{#1}\fi
\ifx \bpublisher  \undefined \def \bpublisher#1{#1}\fi
\ifx \bbtitle  \undefined \def \bbtitle#1{\textit{#1}}\fi
\ifx \bedition  \undefined \def \bedition#1{#1}\fi
\ifx \bseriesno  \undefined \def \bseriesno#1{\textbf{#1}}\fi
\ifx \blocation  \undefined \def \blocation#1{#1}\fi
\ifx \bsertitle  \undefined \def \bsertitle#1{\textit{#1}}\fi
\ifx \bsnm \undefined \def \bsnm#1{#1}\fi
\ifx \bsuffix \undefined \def \bsuffix#1{#1}\fi
\ifx \bparticle \undefined \def \bparticle#1{#1}\fi
\ifx \barticle \undefined \def \barticle#1{}\fi
\ifx \botherref \undefined \def \botherref#1{}\fi
\ifx \url \undefined \def \url#1{\textsf{#1}}\fi
\ifx \bchapter \undefined \def \bchapter#1{}\fi
\ifx \bbook \undefined \def \bbook#1{}\fi
\ifx \bcomment \undefined \def \bcomment#1{#1}\fi
\ifx \oauthor \undefined \def \oauthor#1{#1}\fi
\ifx \citeauthoryear \undefined \def \citeauthoryear#1{#1}\fi
\def \endbibitem {}

\bibitem[\protect\citeauthoryear{{Aly}}{1984}]{Aly:1984}
\begin{barticle}
\bauthor{\bsnm{{Aly}}, \binits{J.J.}}:
\byear{1984},.
\bjtitle{Astrophys. J.}
\bvolume{283},
\bfpage{349}.
\end{barticle}
\endbibitem

\bibitem[\protect\citeauthoryear{{Amari} and {Aly}}{2010}]{Amari:2010}
\begin{barticle}
\bauthor{\bsnm{{Amari}}, \binits{T.}}, \bauthor{\bsnm{{Aly}}, \binits{J.J.}}:
\byear{2010},.
\bjtitle{Astron. Astrophys.}
\bvolume{522},
\bfpage{A52}.
\end{barticle}
\endbibitem

\bibitem[\protect\citeauthoryear{{Amari} \textit{et~al.}}{2013}]{Amari:2013}
\begin{barticle}
\bauthor{\bsnm{{Amari}}, \binits{T.}}, \bauthor{\bsnm{{Aly}}, \binits{J.J.}},
  \bauthor{\bsnm{{Canou}}, \binits{A.}}, \bauthor{\bsnm{{Mikic}}, \binits{Z.}}:
\byear{2013},.
\bjtitle{Astron. Astrophys.}
\bvolume{553},
\bfpage{A43}.
\end{barticle}
\endbibitem

\bibitem[\protect\citeauthoryear{{Aschwanden} and
  {Malanushenko}}{2012}]{Aschwanden:2012}
\begin{botherref}
\oauthor{\bsnm{{Aschwanden}}, \binits{M.J.}}, \oauthor{\bsnm{{Malanushenko}},
  \binits{A.}}:
2012,.
\textit{Solar Phys.}
doi:\doiurl{10.1007/s11207-012-0070-1}.
\end{botherref}
\endbibitem

\bibitem[\protect\citeauthoryear{{Contopoulos}}{2013}]{Contopoulos:2013}
\begin{barticle}
\bauthor{\bsnm{{Contopoulos}}, \binits{I.}}:
\byear{2013},.
\bjtitle{Solar Phys.}
\bvolume{282},
\bfpage{419}.
\end{barticle}
\endbibitem

\bibitem[\protect\citeauthoryear{{Fisher} \textit{et~al.}}{1998}]{Fisher:1998}
\begin{barticle}
\bauthor{\bsnm{{Fisher}}, \binits{G.H.}}, \bauthor{\bsnm{{Longcope}},
  \binits{D.W.}}, \bauthor{\bsnm{{Metcalf}}, \binits{T.R.}},
  \bauthor{\bsnm{{h=80Mm}}, \binits{A.A.}}:
\byear{1998},.
\bjtitle{Astrophys. J.}
\bvolume{508},
\bfpage{885}.
\end{barticle}
\endbibitem

\bibitem[\protect\citeauthoryear{{Forbes}}{2000}]{Forbes:2000}
\begin{barticle}
\bauthor{\bsnm{{Forbes}}, \binits{T.G.}}:
\byear{2000},.
\bjtitle{J. Geophys. Res.}
\bvolume{105},
\bfpage{23153}.
\end{barticle}
\endbibitem

\bibitem[\protect\citeauthoryear{{Gary}}{2001}]{Gary:2001}
\begin{barticle}
\bauthor{\bsnm{{Gary}}, \binits{G.A.}}:
\byear{2001},.
\bjtitle{Solar Phys.}
\bvolume{203},
\bfpage{71}.
\end{barticle}
\endbibitem

\bibitem[\protect\citeauthoryear{{Gosain} \textit{et~al.}}{2013}]{Gosain:2013}
\begin{barticle}
\bauthor{\bsnm{{Gosain}}, \binits{S.}}, \bauthor{\bsnm{{h=80Mm}},
  \binits{A.A.}}, \bauthor{\bsnm{{Rudenko}}, \binits{G.V.}},
  \bauthor{\bsnm{{Anfinogentov}}, \binits{S.A.}}:
\byear{2013},.
\bjtitle{Astrophys. J.}
\bvolume{772},
\bfpage{52}.
\end{barticle}
\endbibitem

\bibitem[\protect\citeauthoryear{{He} and {Wang}}{2008}]{He:2008}
\begin{barticle}
\bauthor{\bsnm{{He}}, \binits{H.}}, \bauthor{\bsnm{{Wang}}, \binits{H.}}:
\byear{2008},.
\bjtitle{J. Geophys. Res.}
\bvolume{113},
\bfpage{5}.
\end{barticle}
\endbibitem

\bibitem[\protect\citeauthoryear{{Jiang} and {Feng}}{2012}]{Jiang:2012}
\begin{barticle}
\bauthor{\bsnm{{Jiang}}, \binits{C.}}, \bauthor{\bsnm{{Feng}}, \binits{X.}}:
\byear{2012},.
\bjtitle{Solar Phys.}
\bvolume{281},
\bfpage{621}.
\end{barticle}
\endbibitem

\bibitem[\protect\citeauthoryear{{Judge}, {Habbal}, and
  {Landi}}{2013}]{Judge:2013}
\begin{botherref}
\oauthor{\bsnm{{Judge}}, \binits{P.G.}}, \oauthor{\bsnm{{Habbal}},
  \binits{S.}}, \oauthor{\bsnm{{Landi}}, \binits{E.}}:
2013,
{From Forbidden Coronal Lines to Meaningful Coronal Magnetic Fields}.
\textit{Solar Phys.}
doi:\doiurl{10.1007/s11207-013-0309-5}.
\end{botherref}
\endbibitem

\bibitem[\protect\citeauthoryear{{Lin}, {Penn}, and {Tomczyk}}{2000}]{Lin:2000}
\begin{barticle}
\bauthor{\bsnm{{Lin}}, \binits{H.}}, \bauthor{\bsnm{{Penn}}, \binits{M.J.}},
  \bauthor{\bsnm{{Tomczyk}}, \binits{S.}}:
\byear{2000},.
\bjtitle{Astrophys. J. Lett.}
\bvolume{541},
\bfpage{L83}.
\end{barticle}
\endbibitem

\bibitem[\protect\citeauthoryear{{Liu} and {Lin}}{2008}]{Liu:2008}
\begin{barticle}
\bauthor{\bsnm{{Liu}}, \binits{Y.}}, \bauthor{\bsnm{{Lin}}, \binits{H.}}:
\byear{2008},.
\bjtitle{Astrophys. J.}
\bvolume{680},
\bfpage{1496}.
\end{barticle}
\endbibitem

\bibitem[\protect\citeauthoryear{{Malanushenko}
  \textit{et~al.}}{2012}]{Malanushenko:2012}
\begin{barticle}
\bauthor{\bsnm{{Malanushenko}}, \binits{A.}}, \bauthor{\bsnm{{Schrijver}},
  \binits{C.J.}}, \bauthor{\bsnm{{DeRosa}}, \binits{M.L.}},
  \bauthor{\bsnm{{Wheatland}}, \binits{M.S.}}, \bauthor{\bsnm{{Gilchrist}},
  \binits{S.A.}}:
\byear{2012},.
\bjtitle{Astrophys. J.}
\bvolume{756},
\bfpage{153}.
\end{barticle}
\endbibitem

\bibitem[\protect\citeauthoryear{{h=80Mm} and {Acton}}{2001}]{h=80Mm:2001}
\begin{barticle}
\bauthor{\bsnm{{h=80Mm}}, \binits{A.A.}}, \bauthor{\bsnm{{Acton}},
  \binits{L.W.}}:
\byear{2001},.
\bjtitle{Astrophys. J.}
\bvolume{554},
\bfpage{416}.
\end{barticle}
\endbibitem

\bibitem[\protect\citeauthoryear{{h=80Mm}, {Canfield}, and
  {McClymont}}{1997}]{h=80Mm:1997}
\begin{barticle}
\bauthor{\bsnm{{h=80Mm}}, \binits{A.A.}}, \bauthor{\bsnm{{Canfield}},
  \binits{R.C.}}, \bauthor{\bsnm{{McClymont}}, \binits{A.N.}}:
\byear{1997},.
\bjtitle{Astrophys. J.}
\bvolume{481},
\bfpage{973}.
\end{barticle}
\endbibitem

\bibitem[\protect\citeauthoryear{{R{\'e}gnier} and
  {Priest}}{2007a}]{Regnier:2007}
\begin{barticle}
\bauthor{\bsnm{{R{\'e}gnier}}, \binits{S.}}, \bauthor{\bsnm{{Priest}},
  \binits{E.R.}}:
\byear{2007}a,.
\bjtitle{Astrophys. J. Lett.}
\bvolume{669},
\bfpage{L53}.
\end{barticle}
\endbibitem

\bibitem[\protect\citeauthoryear{{R{\'e}gnier} and
  {Priest}}{2007b}]{Regnier:2007b}
\begin{barticle}
\bauthor{\bsnm{{R{\'e}gnier}}, \binits{S.}}, \bauthor{\bsnm{{Priest}},
  \binits{E.R.}}:
\byear{2007}b,.
\bjtitle{Astron. Astrophys.}
\bvolume{468},
\bfpage{701}.
\end{barticle}
\endbibitem

\bibitem[\protect\citeauthoryear{{Schrijver}}{2009}]{Schrijver:2009}
\begin{barticle}
\bauthor{\bsnm{{Schrijver}}, \binits{C.J.}}:
\byear{2009},.
\bjtitle{Adv. Space Res.}
\bvolume{43},
\bfpage{739}.
\end{barticle}
\endbibitem

\bibitem[\protect\citeauthoryear{{Schrijver}
  \textit{et~al.}}{2006}]{Schrijver:2006}
\begin{barticle}
\bauthor{\bsnm{{Schrijver}}, \binits{C.J.}}, \bauthor{\bsnm{{De Rosa}},
  \binits{M.L.}}, \bauthor{\bsnm{{Metcalf}}, \binits{T.R.}},
  \bauthor{\bsnm{{Liu}}, \binits{Y.}}, \bauthor{\bsnm{{McTiernan}},
  \binits{J.}}, \bauthor{\bsnm{{R{\'e}gnier}}, \binits{S.}},
  \bauthor{\bsnm{{Valori}}, \binits{G.}}, \bauthor{\bsnm{{Wheatland}},
  \binits{M.S.}}, \bauthor{\bsnm{{Wiegelmann}}, \binits{T.}}:
\byear{2006},.
\bjtitle{Solar Phys.}
\bvolume{235},
\bfpage{161}.
\end{barticle}
\endbibitem

\bibitem[\protect\citeauthoryear{{Song} \textit{et~al.}}{2007}]{Song:2007}
\begin{barticle}
\bauthor{\bsnm{{Song}}, \binits{M.T.}}, \bauthor{\bsnm{{Fang}}, \binits{C.}},
  \bauthor{\bsnm{{Zhang}}, \binits{H.Q.}}, \bauthor{\bsnm{{Tang}},
  \binits{Y.H.}}, \bauthor{\bsnm{{Wu}}, \binits{S.T.}},
  \bauthor{\bsnm{{Zhang}}, \binits{Y.A.}}:
\byear{2007},.
\bjtitle{Astrophys. J.}
\bvolume{666},
\bfpage{491}.
\end{barticle}
\endbibitem

\bibitem[\protect\citeauthoryear{{Tadesse}, {Wiegelmann}, and
  {Inhester}}{2009}]{Tadesse:2009}
\begin{barticle}
\bauthor{\bsnm{{Tadesse}}, \binits{T.}}, \bauthor{\bsnm{{Wiegelmann}},
  \binits{T.}}, \bauthor{\bsnm{{Inhester}}, \binits{B.}}:
\byear{2009},.
\bjtitle{Astron. Astrophys.}
\bvolume{508},
\bfpage{421}.
\end{barticle}
\endbibitem

\bibitem[\protect\citeauthoryear{{Tadesse}
  \textit{et~al.}}{2011}]{Tadesse:2011}
\begin{barticle}
\bauthor{\bsnm{{Tadesse}}, \binits{T.}}, \bauthor{\bsnm{{Wiegelmann}},
  \binits{T.}}, \bauthor{\bsnm{{Inhester}}, \binits{B.}},
  \bauthor{\bsnm{{h=80Mm}}, \binits{A.}}:
\byear{2011},.
\bjtitle{Astron. Astrophys.}
\bvolume{527},
\bfpage{A30}.
\end{barticle}
\endbibitem

\bibitem[\protect\citeauthoryear{{Tadesse}
  \textit{et~al.}}{2012a}]{Tadesse:2012}
\begin{barticle}
\bauthor{\bsnm{{Tadesse}}, \binits{T.}}, \bauthor{\bsnm{{Wiegelmann}},
  \binits{T.}}, \bauthor{\bsnm{{Inhester}}, \binits{B.}},
  \bauthor{\bsnm{{h=80Mm}}, \binits{A.}}:
\byear{2012}a,.
\bjtitle{Solar Phys.}
\bvolume{281},
\bfpage{53}.
\end{barticle}
\endbibitem

\bibitem[\protect\citeauthoryear{{Tadesse}
  \textit{et~al.}}{2012b}]{Tadesse:2012a}
\begin{barticle}
\bauthor{\bsnm{{Tadesse}}, \binits{T.}}, \bauthor{\bsnm{{Wiegelmann}},
  \binits{T.}}, \bauthor{\bsnm{{Inhester}}, \binits{B.}},
  \bauthor{\bsnm{{h=80Mm}}, \binits{A.}}:
\byear{2012}b,.
\bjtitle{Solar Phys.}
\bvolume{277},
\bfpage{119}.
\end{barticle}
\endbibitem

\bibitem[\protect\citeauthoryear{{Tadesse}
  \textit{et~al.}}{2013a}]{Tadesse:2013}
\begin{barticle}
\bauthor{\bsnm{{Tadesse}}, \binits{T.}}, \bauthor{\bsnm{{Wiegelmann}},
  \binits{T.}}, \bauthor{\bsnm{{Inhester}}, \binits{B.}},
  \bauthor{\bsnm{{MacNeice}}, \binits{P.}}, \bauthor{\bsnm{{h=80Mm}},
  \binits{A.}}, \bauthor{\bsnm{{Sun}}, \binits{X.}}:
\byear{2013}a,.
\bjtitle{Astron. Astrophys.}
\bvolume{550},
\bfpage{A14}.
\end{barticle}
\endbibitem

\bibitem[\protect\citeauthoryear{{Tadesse}
  \textit{et~al.}}{2013b}]{Tadesse:2013ad}
\begin{barticle}
\bauthor{\bsnm{{Tadesse}}, \binits{T.}}, \bauthor{\bsnm{{Wiegelmann}},
  \binits{T.}}, \bauthor{\bsnm{{MacNeice}}, \binits{P.J.}},
  \bauthor{\bsnm{{Olson}}, \binits{K.}}:
\byear{2013}b,.
\bjtitle{Astrophys. and Space Science}
\bvolume{347},
\bfpage{21}.
\end{barticle}
\endbibitem

\bibitem[\protect\citeauthoryear{{Tadesse}
  \textit{et~al.}}{2013c}]{Tadesse:2013sol}
\begin{barticle}
\bauthor{\bsnm{{Tadesse}}, \binits{T.}}, \bauthor{\bsnm{{Wiegelmann}},
  \binits{T.}}, \bauthor{\bsnm{{MacNeice}}, \binits{P.J.}},
  \bauthor{\bsnm{{Olson}}, \binits{K.}}:
\byear{2013}c,.
\bjtitle{Astrophys. and Space Science}
\bvolume{348},
\bfpage{21}.
\end{barticle}
\endbibitem

\bibitem[\protect\citeauthoryear{{Valori}, {Kliem}, and
  {Keppens}}{2005}]{valori:2005}
\begin{barticle}
\bauthor{\bsnm{{Valori}}, \binits{G.}}, \bauthor{\bsnm{{Kliem}}, \binits{B.}},
  \bauthor{\bsnm{{Keppens}}, \binits{R.}}:
\byear{2005},.
\bjtitle{Astron. Astrophys.}
\bvolume{433},
\bfpage{335}.
\end{barticle}
\endbibitem

\bibitem[\protect\citeauthoryear{{Wheatland} and
  {R{\'e}gnier}}{2009}]{Wheatland:2009}
\begin{barticle}
\bauthor{\bsnm{{Wheatland}}, \binits{M.S.}}, \bauthor{\bsnm{{R{\'e}gnier}},
  \binits{S.}}:
\byear{2009},.
\bjtitle{Astrophys. J. Lett.}
\bvolume{700},
\bfpage{L88}.
\end{barticle}
\endbibitem

\bibitem[\protect\citeauthoryear{{Wheatland}, {Sturrock}, and
  {Roumeliotis}}{2000}]{Wheatland:2000}
\begin{barticle}
\bauthor{\bsnm{{Wheatland}}, \binits{M.S.}}, \bauthor{\bsnm{{Sturrock}},
  \binits{P.A.}}, \bauthor{\bsnm{{Roumeliotis}}, \binits{G.}}:
\byear{2000},.
\bjtitle{Astrophys. J.}
\bvolume{540},
\bfpage{1150}.
\end{barticle}
\endbibitem

\bibitem[\protect\citeauthoryear{{Wiegelmann}}{2004}]{Wiegelmann:2004}
\begin{barticle}
\bauthor{\bsnm{{Wiegelmann}}, \binits{T.}}:
\byear{2004},.
\bjtitle{Solar Phys.}
\bvolume{219},
\bfpage{87}.
\end{barticle}
\endbibitem

\bibitem[\protect\citeauthoryear{{Wiegelmann}}{2007}]{Wiegelmann:2007}
\begin{barticle}
\bauthor{\bsnm{{Wiegelmann}}, \binits{T.}}:
\byear{2007},.
\bjtitle{Solar Phys.}
\bvolume{240},
\bfpage{227}.
\end{barticle}
\endbibitem

\bibitem[\protect\citeauthoryear{{Wiegelmann} and
  {Inhester}}{2010}]{Wiegelmann:2010}
\begin{barticle}
\bauthor{\bsnm{{Wiegelmann}}, \binits{T.}}, \bauthor{\bsnm{{Inhester}},
  \binits{B.}}:
\byear{2010},.
\bjtitle{Astron. Astrophys.}
\bvolume{516},
\bfpage{A107}.
\end{barticle}
\endbibitem

\bibitem[\protect\citeauthoryear{{Wiegelmann} and
  {Sakurai}}{2012}]{Wiegelmann:2012W}
\begin{barticle}
\bauthor{\bsnm{{Wiegelmann}}, \binits{T.}}, \bauthor{\bsnm{{Sakurai}},
  \binits{T.}}:
\byear{2012},.
\bjtitle{Living Rev. Solar Phys.}
\bvolume{9},
\bfpage{5}.
\burl{http:// http://solarphysics.livingreviews.org/Articles/lrsp-2012-5/}.
\end{barticle}
\endbibitem

\bibitem[\protect\citeauthoryear{{Wiegelmann}, {Inhester}, and
  {Sakurai}}{2006}]{Wiegelmann:2006}
\begin{barticle}
\bauthor{\bsnm{{Wiegelmann}}, \binits{T.}}, \bauthor{\bsnm{{Inhester}},
  \binits{B.}}, \bauthor{\bsnm{{Sakurai}}, \binits{T.}}:
\byear{2006},.
\bjtitle{Solar Phys.}
\bvolume{233},
\bfpage{215}.
\end{barticle}
\endbibitem

\end{thebibliography}
\end{document}